\documentclass[twocolumn,aps,pra,showpacs,superscriptaddress]{revtex4}
\usepackage{times}
\usepackage{graphicx}
\usepackage{longtable}
\usepackage{color}
\usepackage{amsmath}
\usepackage{amssymb}

\newcommand{\RN}{RN}

\newcommand{\ket}[1]{\vert{#1}\rangle}
\newcommand{\bra}[1]{\langle{#1}\vert}

\newcommand{\beq}{\begin{equation}}
\newcommand{\eeq}{\end{equation}}

\newcommand{\beqa}{\begin{eqnarray}}
\newcommand{\eeqa}{\end{eqnarray}}

\def\ket#1{|#1\rangle}
\def\bra#1{\langle #1|}

\def\opone{\leavevmode\hbox{\small1\kern-3.8pt\normalsize1}}

\def\endproof{\vrule height6pt width6pt depth0pt}

\begin{document}

\title{Bell scenarios in which nonlocality and entanglement are inversely related}

%%%%%%%%%%%%%%%%%%%%%%%%%%%%%%%%%%%%%%%%%%%%%%%%%%%%%%%%%%%%%%%%%%%

\author{Giuseppe~Vallone}
\affiliation{Department of Information Engineering, University of Padova, I-35131 Padova, Italy}
\affiliation{Dipartimento di Fisica della ``Sapienza'' Universit\`{a} di Roma, I-00185 Roma, Italy}
\author{Gustavo~Lima}
\affiliation{Center for Optics and Photonics, MSI-Nucleus on Advanced Optics, Departamento de F\'{\i}sica, Universidad de Concepci\'{o}n, 160-C Concepci\'{o}n, Chile}
\author{Esteban~S.~G\'omez}
\affiliation{Center for Optics and Photonics, MSI-Nucleus on Advanced Optics, Departamento de F\'{\i}sica, Universidad de Concepci\'{o}n, 160-C Concepci\'{o}n, Chile}
\author{Gustavo~Ca\~nas}
\affiliation{Center for Optics and Photonics, MSI-Nucleus on Advanced Optics, Departamento de F\'{\i}sica, Universidad de Concepci\'{o}n, 160-C Concepci\'{o}n, Chile}
\author{Jan-{\AA}ke~Larsson}
 \affiliation{Institutionen f\"or Systemteknik, Link\"opings Universitet, SE-58183 Link\"oping, Sweden}
\author{Paolo~Mataloni}
 \affiliation{Dipartimento di Fisica della ``Sapienza'' Universit\`{a} di Roma, I-00185 Roma, Italy}
 \affiliation{Istituto Nazionale di Ottica (INO-CNR), L.go E. Fermi 6, I-50125 Florence, Italy}
\author{Ad\'an~Cabello}
 %\email{adan@us.es}
 \affiliation{Departamento de F\'{\i}sica Aplicada II, Universidad de
 Sevilla, E-41012 Sevilla, Spain}

%%%%%%%%%%%%%%%%%%%%%%%%%%%%%%%%%%%%%%%%%%%%%%%%%%%%%%%%%%%%%%%%%%%

\date{\today}

%First version: April 2011 (Roma).
%This version: July 11, 2013 (Benasque).

%%%%%%%%%%%%%%%%%%%%%%%%%%%%%%%%%%%%%%%%%%%%%%%%%%%%%%%%%%%%%%%%%%%

\begin{abstract}
We show that for two-qubit chained Bell inequalities with an arbitrary number of measurement settings, nonlocality and entanglement are not only different properties but are inversely related. Specifically, we analytically prove that in absence of noise, robustness of nonlocality, defined as the maximum fraction of detection events that can be lost such that the remaining ones still do not admit a local model, and concurrence are inversely related for any chained Bell inequality with an arbitrary number of settings. The closer quantum states are to product states, the harder it is to reproduce quantum correlations with local models. We also show that, in presence of noise, nonlocality and entanglement are simultaneously maximized only when the noise level is equal to the maximum level tolerated by the inequality; in any other case, a more nonlocal state is always obtained by reducing the entanglement. {In addition, we observed that robustness of nonlocality and concurrence are also inversely related 
for the Bell scenarios defined by the tight two-qubit three-setting $I_{3322}$ inequality, and the tight two-qutrit inequality $I_3$.}
\end{abstract}

%%%%%%%%%%%%%%%%%%%%%%%%%%%%%%%%%%%%%%%%%%%%%%%%%%%%%%%%%%%%%%%%%%%

%%Entanglement and quantum nonlocality (e.g. EPR paradox, Bell's inequalities, GHZ states, etc.)
%%Entanglement production and manipulation

\maketitle

%%%%%%%%%%%%%%%%%%%%%%%%%%%%%%%%%%%%%%%%%%%%%%%%%%%%%%%%%%%%%%%%%%%

\section{Introduction}

%%%%%%%%%%%%%%%%%%%%%%%%%%%%%%%%%%%%%%%%%%%%%%%%%%%%%%%%%%%%%%%%%%%

Nonlocality and entanglement are two core concepts in quantum information. If $p_\rho(ab)$ is the joint probability that Alice obtains $a=1$ and Bob $b=1$ on a system prepared in state $\rho$, nonlocality is the impossibility of expressing $p_\rho(ab)$ as $\sum_{\lambda}p_\rho(\lambda)p_\rho(a,\lambda)p_\rho(b,\lambda)$, where $\lambda$ are preestablished classical correlations \cite{bell64ph}. Entanglement is the impossibility of expressing a quantum state as a convex combination of separable states. Nonlocality and entanglement are related concepts in the sense that, to have nonlocality, entanglement is needed \cite{gisin91pla}. The difference between both concepts has been pointed out before.
{ First, it was noticed that there are entangled
states which do not violate specific Bell inequalities \cite{wern00pra}.
Then, in Ref. \cite{acin05prl}, the statistical strength of Bell tests was studied,
showing that stronger tests (for a given family of Bell
inequalities) require nonmaximally entangled states.
Similarly, it was shown in \cite{acin02pra} that
nonmaximally entangled states allow for larger
violations (or equivalently a stronger resistance to noise) of the $I_3$ two-qutrit inequality \cite{CGLMS02}.
In \cite{junge10ar} it was demonstrated
that, for general bipartite Bell inequalities with $n$ inputs, $n$
outputs, and $n$-dimensional Hilbert spaces, the entropy of entanglement
of the state is essentially irrelevant in obtaining
large violation.
Finally, in \cite{vidi10ar,lian11pra}, it is shown that, for certain inequalities, weakly entangled
states outperform maximally entangled ones of arbitrary dimension.}

One difficulty in reaching a general conclusion about the relationship between nonlocality and entanglement is that of finding a general scenario where incontrovertible measures of nonlocality and entanglement can be compared. Bipartite scenarios have the advantage that any of the many measures of entanglement assign zero entanglement to product states and maximum entanglement to maximally entangled states \cite{woot98prl,IconcPRA}. Nonlocality is a more delicate issue since different restrictions on the number of measurement settings usually lead to different measures of nonlocality. This suggests that to study such relationship, one needs to consider a general scenario in which each party can perform an arbitrary number of local measurements.

The structure of the paper is the following: In Sec.~\ref{Sec2} we define a measure of nonlocality called robustness of nonlocality that will be used through all the paper. In Sec.~\ref{Sec3} we discuss a general bipartite scenario in which both parties have the same number of settings and prove that, no matter the number of settings, robustness of nonlocality and entanglement are inversely related. We then study how noise affects this conclusion. In Sec.~\ref{Sec4} we numerically explore 
the second simplest tight bipartite Bell inequality $I_{3322}$~\cite{CG04},  
which has three settings per party, each with two outcomes. In Sec.~\ref{Sec5}, 
we study a tight two-qutrit Bell inequality $I_3$~\cite{CGLMS02}. In all cases considered we observe the same behavior, namely, that entanglement and robustness of nonlocality are inversely related.

%%%%%%%%%%%%%%%%%%%%%%%%%%%%%%%%%%%%%%%%%%%%%%%%%%%%%%%%%%%%%%%%%%%

\section{Robustness of nonlocality}
\label{Sec2}

%%%%%%%%%%%%%%%%%%%%%%%%%%%%%%%%%%%%%%%%%%%%%%%%%%%%%%%%%%%%%%%%%%%

For an ensemble of entangled particles in a state $\ket{\psi}$ and a given Bell inequality, we define the {\it robustness of nonlocality} (\RN) against loss of local information as the maximum fraction of random particles per observer that can be lost such that the remaining ones can violate the Bell inequality. The robustness of nonlocality is related to the minimum detection efficiency, $\eta_{\rm crit}$, required for a loophole-free violation of the Bell inequality \cite{pear70prd} as \RN$\equiv1-\eta_{\rm crit}$.

The idea behind this measure of nonlocality is simple: A violation of a Bell inequality with perfect detection efficiency implies that no local model can reproduce the observed joint probabilities. If the minimum detection efficiency is $\eta_{\rm crit}$, this means that no local model exists, even if one locally rejects a fraction \RN\ of the events. Therefore, the larger \RN, the harder it is to reproduce the observed results with local models. Therefore, \RN\ may be taken as a measure of nonlocality. As a measure of entanglement we will use the {concurrence {\cite{woot98prl,IconcPRA}}}.

Any bipartite Bell inequality involving $m_A$ and $m_B$ dichotomic ($\pm1$) observables $A_j$ and $B_k$ on Alice's and Bob's sides,
respectively, can be written in the following form:
\beq
\langle\mathcal S\rangle_\rho\leq S_{\text{LHV}}\,,
\eeq
where  $\langle\mathcal S\rangle_\rho$ is the expectation value of $\mathcal S$ in the state
$\rho$ and
\beq\label{general}
\mathcal S=\sum^{m_A}_{j=1}\sum^{m_A}_{k=1}c_{jk}p(a_jb_k)+\sum^{m_A}_{j=1}\alpha_{j}p(a_j)+
\sum^{m_B}_{k=1}\beta_{k}p(b_k)\,.
\eeq
In the previous expression, $p(a_jb_j)=p(A_j=1,B_k=1)$ are the joint probabilities of detecting
 the $+1$ eigenvectors $\ket{a_j}$ and $\ket{b_k}$ of the observables $A_j$ and $B_k$.
 If the observables have $d_A$ and $d_B$ outcomes, any Bell inequality can be 
 expressed in a similar way by using only the first $d_A-1$ and $d_B-1$ outcomes.

Let us now evaluate the effect of detection inefficiency.
For inequalities involving only $+1$ outcomes such as \eqref{general}, it is customary to assume that no-detection events
do not contribute to the inequality (they can be seen as detection on the ``$-1$'' outcome). 
However, in order to compute the robustness of non-locality, it is necessary to optimize over all possible strategies  
for the no-detection events; for instance, whenever Alice does not get a detection, she can choose to always  output
 $+1$ for observables $A_1$ and output $-1$ for all other observables \cite{brun08pla}.
 { It is worth noting that, instead of grouping inconclusive
events with one of the outcomes, different strategies can be
used. For instance, a further outcome, corresponding to nondetections,
can be added to the observables, \cite{mass02pra}, or one
can also choose to treat nondetections as simply ``undefined'' \cite{lars98pra}.
However, these strategies will require a modification of
the Bell inequality. In the present paper we will study the robustness
of nonlocality by assigning one of the observable
outcomes to inconclusive events.
} 

Each strategy giving
a definite output to each observable
is completely equivalent to relabeling the inputs or outputs
of a Bell inequality and using the ``$-1$'' outcome in the case
of no-detection for any observable.
 To give an example,  the inequality \eqref{general} with Alice giving output $+1$ only for observable $A_1$ in the case of no detection
 is equivalent to replacing $p(a_1b_k)\rightarrow p(b_k)-p(a_1b_k)$ and $p(a_1)\rightarrow 1-p(a_1)$ and using the $-1$ outcome in 
 the case of a no-detection event for any observable. 
 
% From the experimental point of view, it is easy to deal with no-detection events corresponding to $-1$ outcomes:
% for instance, in many experimental realization with continuous pump source, 
% the entangled photon pair is produced probabilistically. 
% Therefore, Alice and Bob cannot know when the pair
%is produced and whether their photons are not detected because of detection
%inefficiency or because the pair was not produced.
% Interpreting the no-detection events as $-1$ outcomes implies that
% only the true detections contribute to the inequality while no-detection events do not.
  
  It can also be noted that from the experimental viewpoint, assigning
$-1$ outcomes for non-detection gives a simple way to handle these events.
This is because with this assignment, no-detection events do not
contribute to the inequality, so that there is no need to distinguish
whether there was a pair produced but no detection, or if there was no
pair produced. Distinguishing these are sometimes nontrivial, for
example in a continuously pumped experiment, but this is not needed with
the suggested assignment. 

 Thus, since any no-detection strategy is equivalent to rewrite the Bell inequality, the robustness of nonlocality RN can
 be evaluated by  optimizing over all possible ways of rewriting the inequality and using the $-1$ outcome in the case of non detection
 (in the case of observable with $d$ outcomes, the last outcome is typically used in the case of non-detection).
 In order to violate a Bell inequality written as \eqref{general}, 
in the case of detection efficiencies $\eta_A$ and $\eta_B$, the following relation must hold:
 \beq
\begin{split}
\eta_A\eta_B\sum^{m_A}_{j=1}\sum^{m_A}_{k=1}c_{jk}p(a_jb_k)+\eta_A\sum^{m_A}_{j=1}\alpha_{j}p(a_j)+
\\
+\eta_B\sum^{m_B}_{k=1}\beta_{k}p(b_k)>S_{\text{LHV}}\,.
\end{split}
\eeq
Eberhard first showed that states with lower entanglement  allow a violation of the
Clauser-Horne-Simony-Holt (CHSH) inequality \cite{clau69prl} with lower required detection efficiency \cite{eber93pra}
with respect to maximally entangled states. 
{Low entangled
states tolerate smaller efficiencies when one of the two particles
is always detected \cite{cabe07prl,brun07prl}. The same occurs in the $n$-site Clauser-Horne
inequality \cite{lars01pra}.
  In \cite{vert10prl}, it was noticed that nonmaximally entangled
states of two qudits can lower the required detection
efficiency with respect to maximally entangled states.
}
Recently, it was shown
that states with low entanglement can be also useful for EPR-steering with low detection inefficiencies \cite{wise07prl,vall13pra}.

In the following sections we will demonstrate that states with low entanglement can tolerate lower detection inefficiency for the violation
of different Bell inequalities. In particular, we will show that 
  the robustness of non-locality RN and the entanglement are inversely correlated for the studied inequalities.

%%%%%%%%%%%%%%%%%%%%%%%%%%%%%%%%%%%%%%%%%%%%%%%%%%%%%%%%%%%%%%%%%%

\section{Robustness of nonlocality vs concurrence for chained Bell inequalities}
 \label{Sec3}

%%%%%%%%%%%%%%%%%%%%%%%%%%%%%%%%%%%%%%%%%%%%%%%%%%%%%%%%%%%%%%%%%%%

{Pearle \cite{pear70prd} and  Braunstein and Caves (BC) \cite{brau89klu, brau90ap}} 
introduced a generalization of the CHSH \cite{clau69prl} and Clauser-Horne (CH) \cite{clau74prd}
Bell inequalities, known as chained Bell inequalities, in which
Alice and Bob choose among $M \ge 2$ settings. Chained Bell
inequalities have some interesting applications: The case $M=3$
fixes a loophole that occurs in some experiments based on the
CHSH inequality \cite{aerts99prl}. Besides, it reduces the
number of trials needed to rule out local hidden variable
theories \cite{peres00fp}, and improves the security of some
quantum key distribution protocols \cite{barr05prl}. In the
case in which $M$ tends to infinity, the inequality allows one
to discard nonlocal hidden variable theories with a nonzero
local fraction \cite{barr06prl}. Chained Bell inequalities have
been experimentally tested using pairs of photons, with $M=3$
\cite{vall11pra}, $4$ \cite{bosc97prl}, and $21$
\cite{barb05pla}. {It was recently shown than they can be used for
randomness expansion \cite{dhar13pra}.}

The version of the chained Bell inequalities introduced in \cite{bosc97prl}, which is symmetric under the permutation of
Alice and Bob, reads (by using the notation of \eqref{general})
\begin{equation}
\langle S_M\rangle_\rho\leq0,
 \label{BCH}
\end{equation}
where
%\begin{equation}
 \begin{align}
 S_M = &p(a_{M}b_{M})+\sum^{M}_{k=2}\left[p(a_kb_{k-1})+p(a_{k-1}b_k)\right] \nonumber \\
 &- p(a_1b_1)-\sum^{M}_{k=2}\left[p(a_k)+p(b_k)\right],
 \label{SBCH}
 \end{align}
%\end{equation}
%and $(S_M)_\rho$ is the expectation value of $S_M$ in the state
%$\rho$. 

%Here, $a_k$ ($b_k$) with $k=1,\ldots,M$ represents
%dichotomic {\color{red}($\pm1$)} observables, and $p(a_{k}b_{k})$ is the joint
%probability of obtaining $a_k=b_k=1$.

The minimum detection efficiency required for a loophole-free
violation of chained Bell inequalities for any $M \ge 2$ using
maximally entangled states has been obtained in
\cite{cabe09pra}. The fact that the maximum quantum violation
of chained Bell inequalities is always achieved with maximally
entangled states \cite{wehn06pra} might suggest that the
minimum detection efficiency occurs for maximally entangled
states, but no proof exists of whether the detection efficiency
for the chained Bell inequalities can indeed be reduced when
one considers more general classes of entangled states. Indeed,
for case $M=2$, corresponding to the CH inequality (that is equivalent the CHSH), the minimum
detection efficiency occurs for almost product states
\cite{eber93pra, lars01pra}.

In the following we will show that, in absence of noise (e.g., considering pure states),
the states with higher robustness of nonlocality (or the minimum detection efficiency) for any chained Bell
inequality written in the form of \eqref{SBCH} are \emph{almost product states} for which
the robustness of nonlocality tends to
\begin{equation}\label{RNbc}
\mathrm{RN}_{M}=\frac{1}{2M-1}.
\end{equation}
The important point here is that this value is \emph{larger}
than the maximum value of {\RN$_M$}
for maximally entangled states \cite{cabe09pra}, namely,
\begin{equation}
\text{\RN$_M$}^{\text{MES}}=\frac{M\cos\left(\frac{\pi}{2M}\right)-M+1}{M\cos\left(\frac{\pi}{2M}\right)+M-1}.
\end{equation}
Moreover, for Bell inequalities of the form \eqref{BCH} with fixed $M$, we will show that the
value in \eqref{RNbc} is the maximum achievable robustness of nonlocality for any quantum state.
This shows that, for all chained Bell inequalities, entanglement and nonlocality of pure states are inversely related.

{\em Theorem:} The maximum of the robustness of nonlocality of inequality \eqref{BCH} is $RN_M=\frac{1}{2M-1}$
and can be obtained by almost product state.

{\em Proof:} Assuming the same detection efficiency for every party and
setting, i.e., $\eta_A=\eta_B=\eta$, the value of $S_M$ becomes
\begin{equation}
\eta^2(S_M)_\rho-\eta(1-\eta)\sum^{M}_{k=2}[p_\rho(a_k)+p_\rho(b_k)],
\end{equation}
where $p_\rho(a_k)$ is the expectation value of $p(a_k)$ in the
state $\rho$. Therefore, inequality \eqref{BCH} is violated
when $\eta > \eta^{(M)}_{\rm crit}$, with
\begin{equation}
\eta^{(M)}_{\rm crit} = \frac{\sum^{M}_{k=2}\left[p_\rho(a_k)+p_\rho(b_k)\right]}
{(S_M)_\rho+\sum^{M}_{k=2}\left[p_\rho(a_k)+p_\rho(b_k)\right]}.
\end{equation}
Since $p_\rho(a_1b_1)\geq 0$, it is easy to show that $(S_M)_\rho +\sum^{M}_{k=2}\left[p_\rho(a_k)+p_\rho(b_k)\right]\leq p_\rho(a_{M}b_{M})+\sum_{k=2}^{M}\left[p_\rho(a_kb_{k-1})+p_\rho(a_{k-1}b_k)\right]$. Then,
 \begin{equation}
 \eta^{(M)}_{\rm crit} \geq
 \frac{\sum_{k=2}^{M}\left[p_\rho(a_k)+p_\rho(b_k)\right]}
 {p_\rho(a_{M}b_{M})+\sum_{k=2}^{M}\left[p_\rho(a_kb_{k-1})+p_\rho(a_{k-1}b_k)\right]}.
 \label{eq:9b}
 \end{equation}
Clearly,
\begin{equation}
 0 \le p_\rho(a_jb_k) \le \min \left[p_\rho(a_j),\, p_\rho(b_k)\right],
\end{equation}
and the lowest possible bound of the right-hand side of \eqref{eq:9b} is obtained when
 $p_\psi(a_jb_k)=p_\psi(a_j)=p_\psi(b_k)$
for $j$ and $k$ not both equal to 1. We obtain
\begin{equation}
\eta^{(M)}_{\rm crit}\geq\frac{(2M-2)p_\psi(a_1)}{(2M-1)p_\psi(a_1)}=\frac{2M-2}{2M-1},
\end{equation}
which cannot be achieved exactly, but arbitrarily close by the following procedure:
Any generic two-qubit pure states $\rho = \ket \psi \bra \psi$, can be written (in a suitable basis) as
\begin{equation}
 \label{state}
 \ket\psi=\sin\frac{\gamma}{2}\ket{00}-\cos\frac{\gamma}{2}\ket{11},
\end{equation}
with $0\leq\gamma\leq\pi/2$.
Let us consider the following eigenstates:
\begin{subequations}
\begin{align}
\ket{a_1}=\ket{b_1}&=\cos\frac\theta2\ket{0}+\sin\frac\theta2\ket{1},
\\
\ket{a_k}=\ket{b_k}&=\ket{0},\;\text{with}\;k=2,\dots, M,
\end{align}
\end{subequations}
and choose $\theta$ such that
$\tan^2\frac\theta2=\tan\frac{\gamma}{2}$. Then,
$p_\rho(a_1b_1)=0$ and the critical efficiency becomes
\begin{equation}
\label{best-eta}
\eta^{(M)}_{\rm crit}=\frac{2M-2}{2M-3+\frac{2}{1+\tan\gamma/2}},
%\eta^{(M)}_{\rm crit}=\frac{2M-2}{2M-3+\cos^2\frac\theta2},
\end{equation}
which, when $\gamma$ tends to zero (i.e., {\em when the state
tends to a product state}), tends to
\begin{equation}
\eta^{(M)}_{\rm crit}\xrightarrow{\gamma\rightarrow0}\frac{2M-2}{2M-1}\quad\Rightarrow\quad
\text{\RN$_M$}\xrightarrow{\gamma\rightarrow0}\frac{1}{2M-1}\,,
 \label{mineta}
\end{equation}
concluding our proof. \hfill\endproof

%%%%%%%%%%%%%%%%%%%%%%%%%%%%%%%%%%%%%%%%%%%%%%%%%%%%%%%%%%%%%%%%%%%
% Fig. 1
%%%%%%%%%%%%%%%%%%%%%%%%%%%%%%%%%%%%%%%%%%%%%%%%%%%%%%%%%%%%%%%%%%%

\begin{figure}[t]
 \centering
 \includegraphics[width=0.4\textwidth]{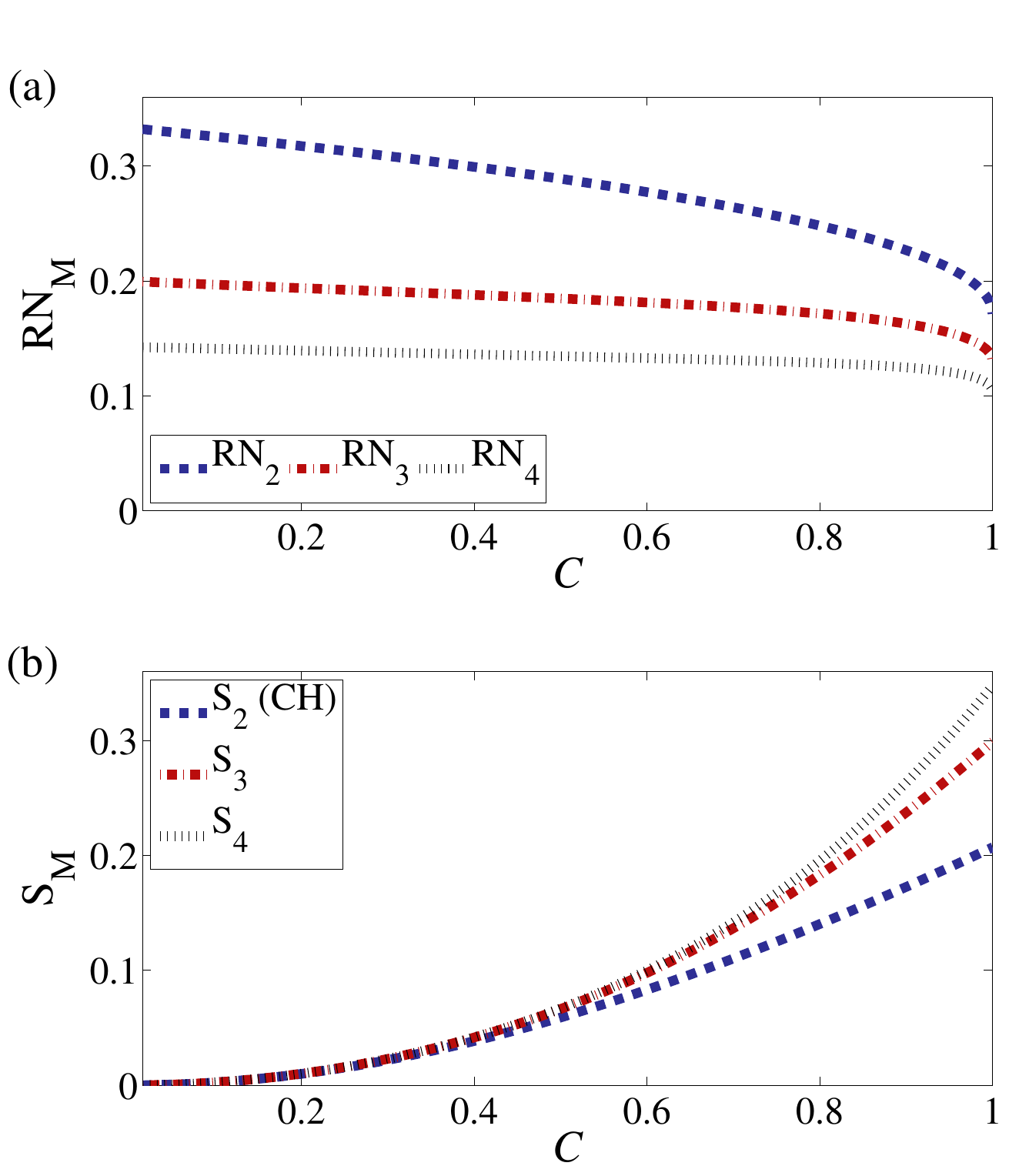}
 \caption{(a) Robustness of nonlocality \RN\ as a function of the concurrence $C$.
 (b) Maximum values of $(S_M)_\rho$ violating the chained Bell inequality as a function of $C$.}
 \label{fig:Etacritk}
\end{figure}

%%%%%%%%%%%%%%%%%%%%%%%%%%%%%%%%%%%%%%%%%%%%%%%%%%%%%%%%%%%%%%%%%%%

We have numerically obtained, by using the method of conjugate gradient, \RN$_M$\ as a function $C$ of the pure state used to violate the inequality and compared it with the corresponding maximal achievable violation of the Bell
inequality $S_M$. Moreover, through exhaustive numerical searches, we have obtained that the   
form \eqref{SBCH} gives the maximum RN for any given state 
(in the specific case of a maximally entangled state this is analytically demonstrated in the Appendix section).
Note that, for nonmaximally entangled states such as \eqref{state}, the concurrence is given by
$C=\sin\gamma$. The results for
$M=2,3,4$ are shown in Fig.~\ref{fig:Etacritk}.
We observe that larger violations of $S_M$ correspond to
lower values of RN$_M$. From Fig.~\ref{fig:Etacritk} one can clearly see see that nonlocality (measured by RN)
and entanglement (measured by $C$) are inversely related: Larger concurrence,
allowing larger violation of the inequality, implies lower RN.

%%%%%%%%%%%%%%%%%%%%%%%%%%%%%%%%%%%%%%%%%%%%%%%%%%%%%%%%%%%%%%%%%%%
\subsection{Adding noise}
%%%%%%%%%%%%%%%%%%%%%%%%%%%%%%%%%%%%%%%%%%%%%%%%%%%%%%%%%%%%%%%%%%

\begin{figure}[t]
 \centering
 \includegraphics[width=0.47\textwidth]{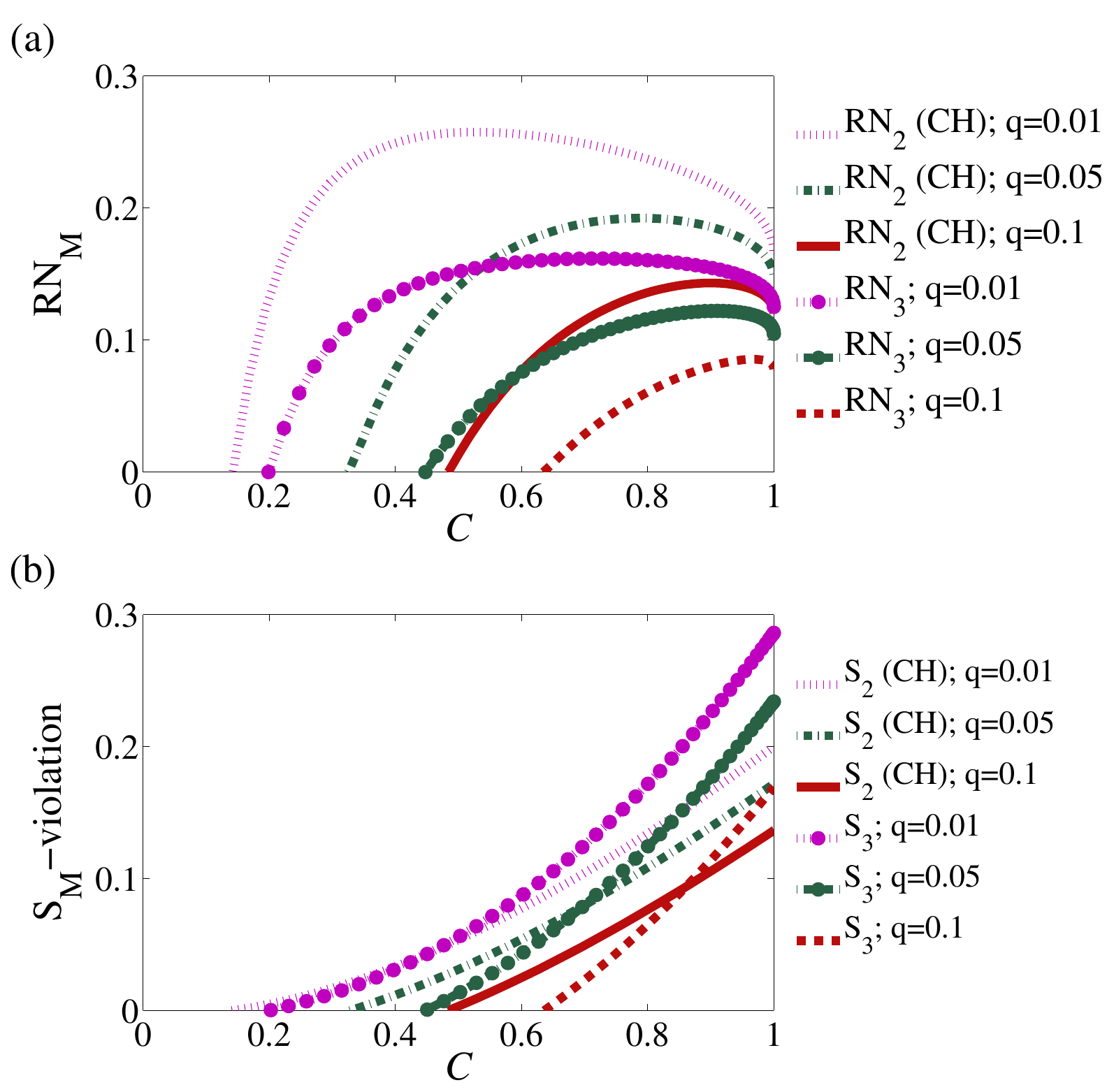}
 \caption{\label{fig:etawithnoise}
 (a) Values of RN$_M$ and (b) maximum violation of the chained Bell inequality for different number of settings and
 different degree of noise ($q$) as a function of the concurrence.}
\end{figure}

How does noise affect this conclusion? In the
presence of white noise, the state becomes
$\rho=(1-q)\ket{\psi}\bra{\psi}+\frac q4\openone$ and the
threshold detection for the chained Bell inequalities
efficiency is changed to
\begin{widetext}
\begin{equation}
 \label{etanoise}
 \eta^{(M)}_{\rm crit}=
 \frac{\sum^{M}_{k=2}\left[p_\rho(a_k)+p_\rho(b_k)\right]+\frac{q}{1-q}(M-1)}
 {(S_M)_\rho+\sum^{M}_{k=2}\left[p_\rho(a_k)+p_\rho(b_k)\right]+\frac{q}{2(1-q)}(M-1)}.
\end{equation}
\end{widetext}
In Fig.~\ref{fig:etawithnoise} we show, for three different
values of noise ($q=0.01$, $q=0.05$, and $q=0.1$), the
dependence of RN$_2$ and RN$_3$ and the maximum values of
$S_2$ and $S_3$ with the degree of entanglement of the initial
pure state. We observe that, when the noise is different from
$0$, the best quantum state giving the lowest threshold is not
an almost separable state, but a nonmaximally entangled state
depending on $q$. However, the lower the noise $q$, the smaller
the entanglement required to obtain the optimal threshold.

%%%%%%%%%%%%%%%%%%%%%%%%%%%%%%%%%%%%%%%%%%%%%%%%%%%%%%%%%%%%%%%%%%%
% Fig. 2
%%%%%%%%%%%%%%%%%%%%%%%%%%%%%%%%%%%%%%%%%%%%%%%%%%%%%%%%%%%%%%%%%%%

%%%%%%%%%%%%%%%%%%%%%%%%%%%%%%%%%%%%%%%%%%%%%%%%%%%%%%%%%%%%%%%%%%%

Furthermore, in Fig.~\ref{fig:etawithnoise}(b) we observe that,
the lower $M$ is, the more resistant to noise is the violation
of the Bell inequality. In fact, it is possible to calculate
the maximum tolerated noise to violate the chained Bell
inequalities. Given $\gamma$ and the maximal violation of
$S_M$ defined as $s^{\text{max}}_M(\gamma)$, the maximum
tolerated noise is
$q_{\text{max}}=\frac{2s^{\text{max}}_M(\gamma)}{2s^{\text{max}}_M(\gamma)+M-1}$.

\begin{figure}[t]
 \centering
 \includegraphics[width=0.45\textwidth]{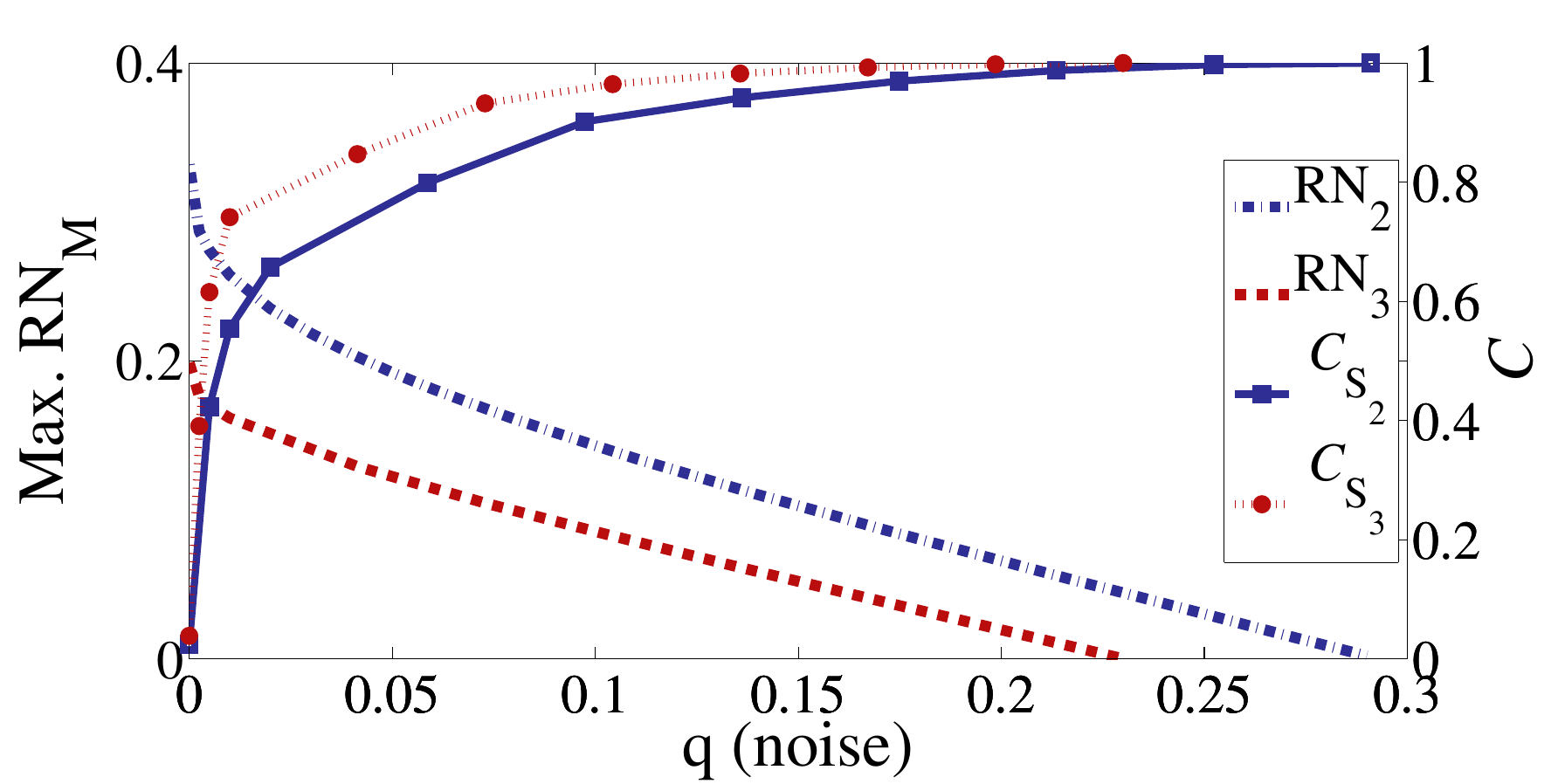}
 \caption{\label{fig:etavsq} Maximal value of RN$_M$ in the presence of noise for the
 $S_2$ and $S_3$ Bell inequalities. The value of $C$
 giving the best RN is shown for each $q$.}
\end{figure}

Using the method of conjugate gradient to minimize
Eq.~(\ref{etanoise}), it is also possible to obtain the
threshold and the required entanglement for any value of the
noise $q$. The results are shown in Fig.~\ref{fig:etavsq}. We
observe that, for chained Bell inequalities, nonlocality and
entanglement are simultaneously maximized {\it only in the case of
extreme noise}, namely the maximum noise level tolerated by the inequality.
A better threshold detection efficiency is
obtained by lowering the noise and suitably decreasing the
entanglement. From this we conclude that nonlocality and
entanglement are synonymous only for extremely noisy scenarios.

%%%%%%%%%%%%%%%%%%%%%%%%%%%%%%%%%%%%%%%%%%%%%%%%%%%%%%%%%%%%%%%%%%%
% Fig. 3
%%%%%%%%%%%%%%%%%%%%%%%%%%%%%%%%%%%%%%%%%%%%%%%%%%%%%%%%%%%%%%%%%%%

%%%%%%%%%%%%%%%%%%%%%%%%%%%%%%%%%%%%%%%%%%%%%%%%%%%%%%%%%%%%%%%%%%%
% Fig. 4
%%%%%%%%%%%%%%%%%%%%%%%%%%%%%%%%%%%%%%%%%%%%%%%%%%%%%%%%%%%%%%%%%%%

%\begin{figure}[t]
% \centering
% \includegraphics[width=0.4\textwidth]{Fig4.pdf}
% \caption{\label{GFig}(a) Robustness of nonlocality \RN\ as a function of the concurrence $C$ and
% (b) maximum violation of the $I_{3322}$ inequality as a function of $C$.}
%\end{figure}

\begin{figure}[t]
 \centering
 \includegraphics[width=0.4\textwidth]{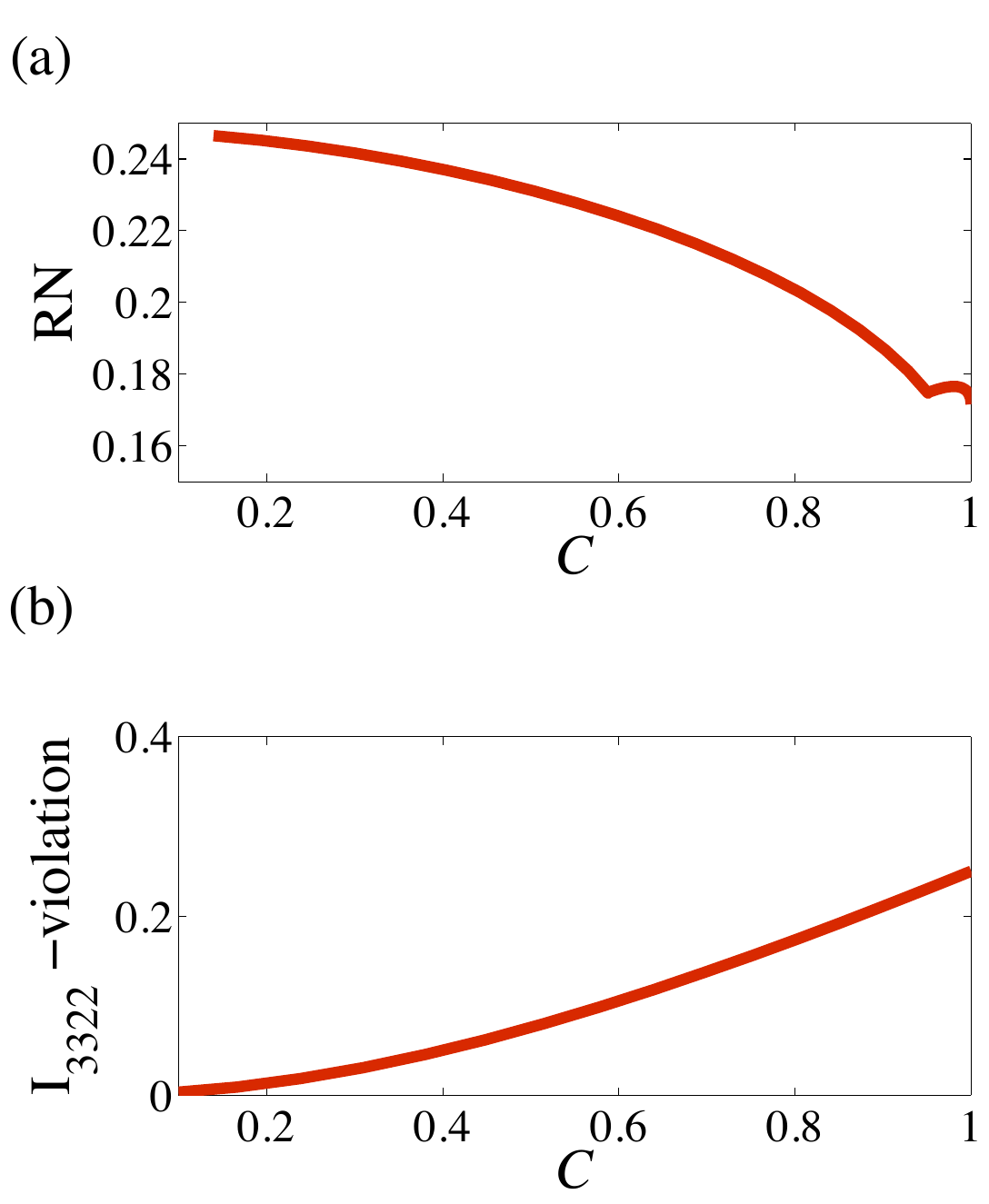}
 \caption{\label{GFig}(a) Robustness of nonlocality \RN\ as a function of the concurrence $C$ and
 (b) maximum violation of the $I_{3322}$ inequality as a function of $C$.}
\end{figure}

%%%%%%%%%%%%%%%%%%%%%%%%%%%%%%%%%%%%%%%%%%%%%%%%%%%%%%%%%%%%%%%%%%%

\section{Robustness of nonlocality vs concurrence for $I_{3322}$ Bell inequality}
\label{Sec4}

%%%%%%%%%%%%%%%%%%%%%%%%%%%%%%%%%%%%%%%%%%%%%%%%%%%%%%%%%%%%%%%%%%%

After the results presented in the previous section, a natural question is whether or not the same behavior occurs for other bipartite Bell inequalities. 
%To check this we have examined the following multi-setting tight two-qubit Bell inequalities: 
%{\color{red}$A_5$, $A_6$, $AII_1$, $AII_2$, $AS_1$, $I_{3322}$, $I_{4422}^{(1)}$,
%$I_{4422}^{(2)}$, $I_{4422}^{(3)}$, $I_{4422}^{(4)}$,
%$I_{4422}^{(6)}$, $I_{4422}^{(8)}$, $I_{4422}^{(9)}$,
%$I_{4422}^{(10)}$, $I_{4422}^{(11)}$, $I_{4422}^{(12)}$,
%$I_{4422}^{(16)}$, $I_{4422}^{(17)}$, $I_{4422}^{(18)}$,
%$I_{4422}^{(20)}$, $S^{51}$, and $S^{52}$, using the notation of Ref.~\cite{brun08pla,Gisin2010}.}
%In all these cases,
%by considering the symmetric and asymmetric versions of each inequality mentioned above,
%the conclusions are similar as those in the previous section. As an example, 
In this section we present the results for the second simplest tight bipartite Bell inequality, namely, $I_{3322}$ \cite{CG04,Froissart81,Sliwa03}, involving three dichotomic measurements on both $A$ and $B$ sides
 (the simplest tight bipartite Bell inequality is the CHSH inequality or $S_2$, studied in the previous section).

The $I_{3322}$ inequality may be written as
\begin{equation}
\begin{aligned}
 \langle I_{3322}\rangle_\rho \leq 0,
\end{aligned}
\end{equation}
where $I_{3322}$ was defined in \cite{CG04} as $I_{3322}=p(a_1b_1)+p(a_1b_2)+p(a_1b_3)+p(a_2b_1)
 +p(a_2b_2)+p(a_3b_1)-p(a_2b_3)-p(a_3b_2)
 -2p(a_1)-p(a_2)-p(b_1)$. However, this form will not lead to the best RN. We have numerically checked that the forms giving the best
 RN are the following: 
%\begin{equation}
% \label{gisin}
%\begin{aligned}
% I^{(1)}_{3322}=&p(a_1b_1)+p(a_1b_2)+p(a_1b_3)+p(a_2b_1)\\
% &+p(a_2b_2)+p(a_3b_1)-p(a_2b_3)-p(a_3b_2)\\
% &-2p(a_1)-p(a_2)-p(b_1)\,,
%\end{aligned}
%\end{equation}
\begin{equation}
 \label{gisin}
\begin{aligned}
 I^{(1)}_{3322}=&p(a_1b_1)-p(a_1b_2)+p(a_1b_3)+p(a_2b_1)\\
 &-p(a_2b_2)+p(a_3b_1)-p(a_2b_3)+p(a_3b_2)\\
 &-p(a_1)-p(a_3)-p(b_1)\,,
\end{aligned}
\end{equation}
that can be obtained from $I_{3322}$ by replacing $p(a_ib_2)\rightarrow p(a_i)-p(a_ib_2)$ and 
\begin{equation}
 \label{gisin}
\begin{aligned}
 I^{(2)}_{3322}=&-p(a_1b_1)-p(a_1b_2)+p(a_1b_3)-p(a_2b_1)\\
 &-p(a_2b_2)-p(a_3b_1)-p(a_2b_3)+p(a_3b_2)\\
 &+p(a_2)+p(b_1)-1\,,
\end{aligned}
\end{equation}
obtained from $I^{(1)}_{3322}$ by $p(a_ib_1)\rightarrow p(a_i)-p(a_ib_1)$ and $p(b_1)\rightarrow 1-p(b_1)$.

Fig.~\ref{GFig} shows the RN and
the violation of the $I_{3322}$ inequality as a function of the
degree of entanglement measured by $C$. We observe
that, in the absence of noise, almost product states are again
those that require lower detection efficiencies. 
For $C\geq0.9507$ the optimal RN is obtained with $I^{(2)}_{3322}$,
while for $C<0.9507$ the optimal RN is obtained with $I^{(1)}_{3322}$.
The optimality of $I^{(2)}_{3322}$ can be analytically shown for maximally entangled states (see the Appendix).

The minimum required efficiency with
maximally entangled states is $0.8284$ as
reported in Ref.~\cite{brun08pla}. Indeed, this value can be obtained analytically. 
For two-qubit systems, the maximum violation of $I_{3322}$ is 1/4, 
and can be achieved with a maximally entangled state, as was previously shown in \cite{CG04}.
Given a maximally entangled
state $\rho'$ that maximally violates $I^{(2)}_{3322}$ and symmetric efficiencies $\eta$, to violate the inequality it is necessary that
%\begin{equation}
% \label{gisin_eta}
%\begin{aligned}
%&\eta^2[p_{\rho'}(a_1b_1)+p_{\rho'}(a_1b_2)+p_{\rho'}(a_1b_3)+p_{\rho'}(a_2b_1)\\
% &+p_{\rho'}(a_2b_2)+p_{\rho'}(a_3b_1)-p_{\rho'}(a_2b_3)-p_{\rho'}(a_3b_2)]\\
% &-\eta[2p_{\rho'}(a_1)+p_{\rho'}(a_2)+p_{\rho'}(b_1)]>0,
%\end{aligned}
%\end{equation}
%or, equivalently,
\begin{equation}
\label{gisin_eta}
\begin{aligned}
&\eta^2 \langle I_{3322}\rangle_{\rho'}+\eta(1-\eta)[p_{\rho'}(a_2)-1]+\\
&\eta(1-\eta)[p_{\rho'}(b_1)-1]-(1-\eta)^2>0.
\end{aligned}
\end{equation}
Remembering that for a maximally entangled state (MES) $p_{\rho'}(a_i)=\frac12$, we obtain
\beq
\eta>2(\sqrt2-1)\simeq0.828\quad\Rightarrow\quad \text{RN}^\text{MES}\simeq0.172.
\eeq
The maximal robustness of non-locality RN$=\frac14$ can be achieved for almost product states.
 If we consider the $I^{(1)}_{3322}$ form of the inequality,
the critical efficiency can be written as $\eta_c=\frac{p_\rho(a_1)+p_\rho(a_3)+p_\rho(b_1)}{\langle I^{(1)}_{3322}\rangle+p_\rho(a_1)+p_\rho(a_3)+p_\rho(b_1)}$. Let us choose $\ket{a_1}=\ket{b_1}=\ket{a_3}=\ket0$, $\ket{a_2}=\cos\frac\theta2\ket0+\sin\frac\theta2\ket1$,
$\ket{b_2}=\ket{b_3}=\sin\frac\theta2\ket0+\cos\frac\theta2\ket1$,
$ \ket\psi=\sin\frac{\gamma}{2}\ket{00}+\cos\frac{\gamma}{2}\ket{11}$. By using $\theta$ such that 
$\cos\theta=\frac{1+\cos\gamma}{2(1+\sin\gamma)}$
we obtain
\beq
\label{best-eta-I3322}
\eta_\text{crit}=
\frac{12 (1+\sin\gamma )}{13+3 \cos\gamma+12 \sin\gamma }\,,
\end{equation}
which, when $\gamma$ tends to zero tends to
\begin{equation}
\eta^{(M)}_{\rm crit}\xrightarrow{\gamma\rightarrow0}\frac{3}{4}\quad\Rightarrow\quad
\text{\RN}\xrightarrow{\gamma\rightarrow0}\frac{1}{4}\,.
 \label{mineta}
\end{equation}
We observe that the maximum RN for the $I_{3322}$ is
greater than the one for the $S_3$ inequality,
which has the same number of
local settings.

%%%%%%%%%%%%%%%%%%%%%%%%%%%%%%%%%%%%%%%%%%%%%%%%%%%%%%%%%%%%%%%%%%%

\section{Robustness of nonlocality vs concurrence for the $I_{3}$ two-qutrit inequality}
\label{Sec5}

%%%%%%%%%%%%%%%%%%%%%%%%%%%%%%%%%%%%%%%%%%%%%%%%%%%%%%%%%%%%%%%%%%%

For the two-qubit Bell inequalities discussed above we have observed that nonlocality and entanglement are inversely related. 
Here we show that this is also true for other bipartite scenarios. For this purpose we repeat our analysis but now for a tight bipartite inequality maximally violated by two-qutrit states, the $I_{3}$ inequality \cite{CGLMS02}.

The inequality is given by
%\begin{eqnarray}
% I_{3}&=&P(A_{1}=B_{1})+P(B_{1}=A_{2}+1)+P(A_{2}=B_{2}) \nonumber \\
% &+&P(B_{2}=A_{1})-P(A_{1}=B_{1}-1)-P(B_{1}=A_{2})\nonumber \\
% &-&P(A_{2}=B_{2}-1)-P(B_{2}=A_{1}-1)\leq 2,
% \end{eqnarray}
$ \widetilde I_3=P(A_{1}=B_{1})+P(B_{1}=A_{2}+1)+P(A_{2}=B_{2}) +P(B_{2}=A_{1})-P(A_{1}=B_{1}-1)-P(B_{1}=A_{2}
 -P(A_{2}=B_{2}-1)-P(B_{2}=A_{1}-1)\leq 2,$
  where
$ P(A_{m}=B_{n}+k)=\sum_{j=1}^{3}P\left(a_{m}^{j}b_{n}^{j+k\,\mathrm{mod}3}\right)$.
Here, $n$ and $m$ ($n,m=1,2$) denote the settings that the parties may choose for the local measurements, and the index $j$ denotes each measurement outcome ($j=1,2,3$). The inequality can be rewritten in the form of \eqref{general}
as $\langle I_{3}\rangle_\rho \leq 0$, with

\begin{widetext}
\beq\label{I3}
\begin{aligned}
 I_{3}=&p(a_1b_1)+p(a_1b_2)+p(a_2b_1)-p(a_2b_2)+
 p(\bar a_1\bar b_1)+p(\bar a_1\bar b_2)+p(\bar a_2\bar b_1)-p(\bar a_2\bar b_2)+
\\
&p(a_1\bar b_1)+p(\bar a_1b_2)+p(\bar a_2b_1)-p(\bar a_2b_2)-p(a_1)-p(\bar a_1)-p(b_1)-p(\bar b_1)\,.
 \end{aligned}
 \eeq
 \end{widetext}
In the previous expression $a_j$ and $b_j$ denote the 1-eigenstates of the observables $A_j$ and $B_k$,
while $\bar a_j$ and $\bar b_j$ denote the 2-eigenstates. Note that, no probability containing the 3-eigenstate is present.
Moreover, $\widetilde I_3=3I_3+2$.

\begin{figure}[t]
 \centering
\includegraphics[width=0.45\textwidth]{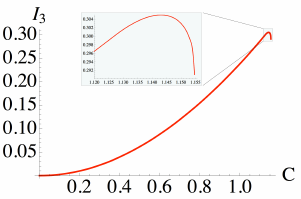}
 \caption{\label{fig:I3} Maximum violation of the $I_{3}$ qutrit inequality as a function of the concurrence $C$. }
\end{figure}

In Fig.~\ref{fig:I3}
we show the maximal achievable violation of $I_3$ in function of the concurrence $C$ while considering the initial two-qutrit state given by 
$\ket{\psi}=\cos\frac{\theta_1}2\cos\frac{\theta_2}2\ket{11}+
\cos\frac{\theta_1}2\sin\frac{\theta_2}2\ket{22}+
\sin\frac{\theta_1}2\ket{33}$,
 where $\theta_k\in[0,\pi]$. The maximally entangled state is obtained when $\theta_2=\pi/2$ and $\theta_1=2\arcsin\frac1{\sqrt3}$.
 The concurrence for the two-qutrit state  described above is given by $C=\sqrt{\sin^2\theta_1+\sin^2\theta_2\cos^4\frac{\theta_1}{2}}$ and
  ranges from 0 to $\frac{2}{\sqrt{3}}$ \cite{IconcPRA}. 
 As was first observed in \cite{acin02pra}, the maximal violation of $I_3$ is obtained with partially entangled states. 
The maximal value achievable with MES is $I^{\text{MES}}_3=\frac2{27} ( 4 \sqrt3-3)\simeq0.291$
 and one can clearly see that it does not correspond to the maximum of $I_3$.

\begin{figure}[t]
 \centering
\includegraphics[width=0.45\textwidth]{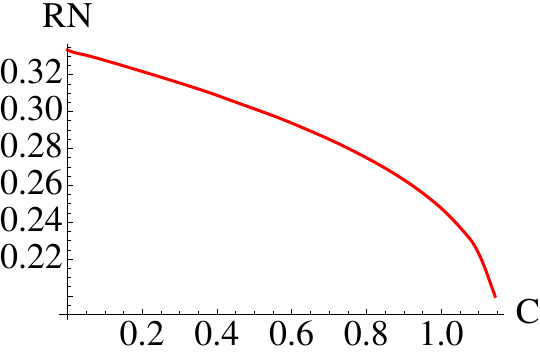}
 \caption{\label{fig:RN_I3} 
RN as a function of the concurrence $C$ for the $I_3$ qutrit inequality.
 }
\end{figure}

We then numerically optimized the robustness of nonlocality RN in a function of $C$, by obtaining the results shown in Fig.~\ref{fig:RN_I3}:
the form of $I_3$ given in \eqref{I3} is the optimal form for maximizing the RN parameter. The optimality
can be analytically shown for maximally entangled states (see the Appendix).
Also in this case RN and entanglement are negatively correlated.

The maximal RN is $\frac13$ and can be obtained for almost product states, as in the previous inequalities.
Let us consider the following entangled state:
\beq
\ket{\psi}=\cos{\frac{\gamma}{2}}\ket{11}+\sin{\frac{\gamma}{2}}\ket{33}\,,
\eeq
and the following measurements
$\ket{a_1}=\ket{b_1}=\ket{3}$,
$\ket{\bar a_1}=\ket{\bar b_1}=\ket{2}$,
$\ket{a_2}=\cos\frac{\theta}{2}\ket1+\sin\frac{\theta}{2}\ket3$,
$\ket{b_2}=\sin\frac{\theta}{2}\ket1+\cos\frac{\theta}{2}\ket3$,
$\ket{\bar a_2}=\ket{\bar b_2}=\ket2$. 
In the form \eqref{I3} the threshold efficiency becomes (non-detection events correspond to $3-$eigenvalues and thus does not contribute to the inequality):
\beq\label{efficiency_I3}
\eta_c=\frac{p_\rho(a_1)+p_\rho(\bar a_1)+p_\rho(b_1)+p_\rho(\bar b_1)}{\langle\widetilde I_3\rangle_\rho+
p_\rho(a_1)+p_\rho(\bar a_1)+p_\rho(b_1)+p_\rho(\bar b_1)}\,.
\eeq

%The quantity appearing in \eqref{efficiency} are the following
%\beq
%\begin{aligned}
%&p(a_1)=p(b_1)=\sin^2\frac{\gamma}{2}
%\\
%&p(\bar a_1)=p(\bar b_1)=0
%\\
%&p(a_1,b_1)=\sin^2\frac{\gamma}{2}
%\\
%&p(a_2,b_1)=\sin^2\frac{\gamma}{2}\sin^2\frac{\theta}{2}
%\\
%&p(a_1,b_2)=\sin^2\frac{\gamma}{2}\cos^2\frac{\theta}{2}
%\\
%&p(a_2,b_2)=\cos^2\frac{\theta}{2}\sin^2\frac{\theta}{2}(1+\sin\gamma)
%\end{aligned}
%\eeq
%while all other bipartite probabilities are zero. 

With the above measurement the critical efficiency becomes
\beq
\eta_c=\frac{4 (1-\cos\gamma)}{3-\cos\theta(2+\cos\theta)-4 \cos\gamma+\sin^2\theta \sin\gamma}\,.
\eeq
If we choose  $\cos\theta=-\frac{1}{\sin\gamma+1}$ and let $\gamma$ go to zero we get
\beq
\eta_c\xrightarrow{\gamma\rightarrow0}\frac23
\quad\Rightarrow\quad
\text{RN}\xrightarrow{\gamma\rightarrow0}\frac13\,.
\eeq

For maximally entangled states, for which $p(a_j)=p(\bar a_j)=p(b_j)=p(\bar b_j)=\frac13$,
the RN is given by
\beq
\text{RN}^{\text{MES}}=\frac{I^{\text{MES}}_3}{I^{\text{MES}}_3+4/3}=
\frac{4\sqrt3-3}{4\sqrt3+15}
\simeq 0.1791\,.
\eeq

%%%%%%%%%%%%%%%%%%%%%%%%%%%%%%%%%%%%%%%%%%%%%%%%%%%%%%%%%%%%%%%%%%%

\section{Conclusions}

%%%%%%%%%%%%%%%%%%%%%%%%%%%%%%%%%%%%%%%%%%%%%%%%%%%%%%%%%%%%%%%%%%%
We would argue that robustness of nonlocality RN is a good measure of nonlocality, since it marks the border where local hidden variable descriptions become possible: The larger robustness of nonlocality is, the harder it is to express the joint probabilities with local models.

We have shown that, for the two-party $M$-setting chained Bell scenario (for any $M\ge 2$ finite), for 
a tight two-qubit Bell inequality $I_{3322}$ and a tight two-qutrit Bell inequality $I_3$, robustness of nonlocality and concurrence, are in the absence of noise, inversely related.

The main result of this paper is the observation that, for many distinct types of Bell scenarios,
larger nonlocality requires smaller entanglement; in the absence of noise, almost product states are the most nonlocal ones. 
We analytically showed that the maximal RN can be achieved with
almost product state. The maximal values of RN (related to the minimum 
required detection efficiency as $\eta_c=1-$RN) are given by RN$=\frac{1}{2M-1}$, RN$=\frac{1}{4}$ and RN$=\frac{1}{3}$ for
the $S_M$ chained Bell inequality, the $I_{3322}$ inequality and the $I_3$ inequality respectively.

When noise is present, the most nonlocal states acquire some amount of entanglement; however, the smaller the noise is, the lower their entanglement becomes.

{
Some questions naturally arise: are the nonlocality and entanglement inversely related  in any
 Bell inequality involving $m_A$, $m_B$ observables with $d_A$ and $d_B$ outcomes? 
If yes, is there some physical mechanism for such counterintuitive behavior? 
These questions require further research. 
}

%%%%%%%%%%%%%%%%%%%%%%%%%%%%%%%%%%%%%%%%%%%%%%%%%%%%%%%%%%%%%%%%%%%

\begin{acknowledgments}
GL, ESG and GC were supported by the CONICYT, AGCI, FONDECYT 1120067, MilenioP10-030-F and PIA-CONICYT PFB0824. GV was supported by
the Strategic-Research-Project QUINTET of the Department
of Information Engineering, University of Padova and the
Strategic-Research-Project QUANTUMFUTURE of the University of Padova.
PM acknowledge the Chistera EU project QUASAR. AC was supported by Project No.\ FIS2011-29400 (MINECO, Spain).
\end{acknowledgments}

%\onecolumngrid

\begin{appendix}
\section{Optimality of detection strategy for maximally entangled states}
In this section we will demonstrate which is the optimal way of rewriting the Bell inequalities analyzed in the main text
in case of maximally entangled states. We start by giving the general framework to solve the optimization.

Let us consider a general bipartite Bell inequality involving $m_A$ and $m_B$
observables $A_j$ and $B_k$ on the Alice and Bob side.
The observables have  $d_A$ and $d_B$ outcomes respectively, $\mu=1, 2,\cdots, d_A$ and $\nu=1, 2,\cdots, d_B$.
Any Bell inequality can be written as \eqref{general}:
\beq
\langle\mathcal S\rangle_\rho\leq S_{\text{LHV}}\,,
\eeq
with
\beq
\begin{split}
\mathcal S=\sum^{m_A}_{j=1}\sum^{m_A}_{k=1}\sum^{d_A-1}_{\mu=1}\sum^{d_A-1}_{\nu=1}c^{\mu\nu}_{jk}p(a^\mu_jb^\mu_k)+
\\
\sum^{m_A}_{j=1}\sum^{d_A-1}_{\mu=1}\alpha^\mu_{j}p(a^\mu_j)+
\sum^{m_B}_{k=1}\sum^{d_B-1}_{\nu=1}\beta^\nu_{k}p(b^\nu_k).
\end{split}
\eeq
In the previous expression $p(a^\mu_jb^\nu_j)=p(A_j=\mu,B_k=\nu)$ are the joint probabilities of detecting
 the $\mu$ and $\nu$ eigenvectors $\ket{a^\mu_j}$ and $\ket{b^\nu_k}$ of the observables $A_j$ and $B_k$.
Note that only the first $d_A-1$ and $d_B-1$ outcomes are involved in the inequality.

When inefficiencies are present it is necessary to give a strategy for the non-detection events. Let us suppose that the strategy on Alice's side
is the following. If Alice is measuring the observable $A_j$ and the particle is not detected, she assigned, with probability $\mathcal A^{(\mu)}_j$,
the outcome $\mu$. Clearly, $\sum^{d_A}_{\mu=1}\mathcal A^{(\mu)}_j=1$. The same happens at Bob's side, with probabilities $\mathcal B^{(\nu)}_k$.
If we consider Alice and Bob inefficiencies as $\eta_A$ and $\eta_B$, the Bell inequality is violated if 
%\begin{widetext}
\beq\label{eq:general_eta}
\begin{split}
\eta_A\eta_B\langle\mathcal S\rangle_\rho+
(1-\eta_A)\eta_BT_A
+
\eta_A(1-\eta_B)T_B+
\\
+(1-\eta_A)(1-\eta_B)
X_{AB}>S_{\text{LHV}}\,,
\end{split}
\eeq
with
\beq
\begin{aligned}
T_A&=\sum_{j,k,\mu,\nu}c^{\mu\nu}_{jk}\mathcal A^{(\mu)}_jp_\rho(b^\nu_k)+
\sum_{j,\mu}\alpha^\mu_{j}\mathcal A^{(\mu)}_j+
\sum_{k,\nu}\beta^\nu_{k}p_\rho(b^\nu_k)
\\
T_B&=\sum_{j,k,\mu,\nu}c^{\mu\nu}_{jk}p_\rho(a^\mu_j)\mathcal B^{(\nu)}_k+
\sum_{j,\mu}\alpha^\mu_{j}p_\rho(a^\mu_j)+
\sum_{k,\nu}\beta^\nu_{k}\mathcal B^{(\nu)}_k
\\
X_{AB}&=\sum_{j,k,\mu,\nu}c^{\mu\nu}_{jk}\mathcal A^{(\mu)}_j\mathcal B^{(\nu)}_k+
\sum_{j,\mu}\alpha^\mu_{j}\mathcal A^{(\mu)}_j+
\sum_{k,\nu}\beta^\nu_{k}\mathcal B^{(\nu)}_k\,.
\end{aligned}
\eeq

%that can be rewritten as
%\beq
%\label{eq:general_eta}
%\begin{aligned}
%\eta_A\eta_B\sum_{j,k,\mu,\nu}c^{\mu\nu}_{jk}\left[p_\rho(a^\mu_jb^\nu_k)-p_\rho(a^\mu_j)\mathcal B^{(\nu)}_k
%-\mathcal A^{(\mu)}_jp_\rho(b^\nu_k)+\mathcal A^{(\mu)}_j\mathcal B^{(\nu)}_k\right]
%+
%\\
%\eta_A
%\Big[\sum_{j,\mu}(\sum_{k,\nu}c^{\mu\nu}_{jk}\mathcal B^{(\nu)}_k+\alpha^\mu_{j})(p_\rho(a^\mu_j)-\mathcal A^{(\mu)}_j)
%\Big]
%+
%\eta_B
%\Big[\sum_{k,\nu}(\sum_{j,\mu}c^{\mu\nu}_{jk}\mathcal A^{(\mu)}_j+\beta^\nu_{k})(p_\rho(b^\nu_k)-\mathcal B^{(\nu)}_k)
%\Big]
%+\gamma
%>0
%\end{aligned}
%\eeq
%\end{widetext}
The sum is taken over $i=1,\ldots,m_A$ and $j=1,\ldots,m_B$ while
 $\mu=1,\ldots,d_A-1$ and $\nu=1,\ldots,d_B-1$: also in the previous expression 
 the outcomes $d_A$ and $d_B$ of each observable are not present.

We start with the chained Bell inequalities, and then analyze the $I_{3322}$ and $I_3$
inequalities.

\subsection{Chained Bell inequalities}
For the chained Bell inequalities of section \ref{Sec3}, we have dichotomic observables.
Then, in the case of non detection on the observable $A_j$, Alice chooses to output the $+1$ outcome with probability $\mathcal A_j$
and the $-1$ outcome with probability $1-\mathcal A_j$. The same happens to Bob.
Remembering that,  for MES, $p_\rho(a_j)=p_\rho(b_k)=\frac12$ we have:
\beq\label{1}
\begin{aligned}
T_A=T_B=&\frac{M-1}{2}
\\
X_{AB}=&\mathcal A_M\mathcal B_M
+\sum^{M}_{k=2}(
\mathcal A_k\mathcal B_{k-1}+\mathcal A_{k-1}\mathcal B_k
)
+
\\
&-\mathcal A_1\mathcal B_1-\sum^{M}_{k=2}(\mathcal A_k+\mathcal B_k)\,.
\end{aligned}
\eeq
Since $X_{AB}$ corresponds to the chained Bell inequality applied to the classical probabilities $\mathcal A_k$ and 
$\mathcal B_k$, we have $X_{AB}\leq 0$.
In order to maximize the Bell parameter it is necessary to choose the $\mathcal A_k$'s and $\mathcal B_k$'s that maximize $X_{AB}$.
The trivial choice $\mathcal A_k=\mathcal B_k=0$, $\forall k$ satisfies this requirement.
It is worth noticing that the choice $\mathcal A_k=\mathcal B_k=0$, $\forall k$, corresponds precisely to consider all non-detections
as $-1$ outputs for the inequality written as \eqref{BCH}.

\subsection{$I_{3322}$ inequality}

Let us consider the $I_{3322}$ inequality written in its original form $I_{3322}=p(a_1b_1)+p(a_1b_2)+p(a_1b_3)+p(a_2b_1)
 +p(a_2b_2)+p(a_3b_1)-p(a_2b_3)-p(a_3b_2)
 -2p(a_1)-p(a_2)-p(b_1)$. In the case of inefficiencies with non maximally entangled states we have
 \beq
 \begin{aligned}
 T_A=&-\frac12(\mathcal A_1+\mathcal A_2+1)
\\
 T_B=&\frac12(\mathcal B_1+\mathcal B_2-3)
\\
X_{AB}=&\mathcal A_3(\mathcal B_1-\mathcal B_2) +\mathcal B_3(\mathcal A_1-\mathcal A_2)+
\mathcal A_1\mathcal B_1+\mathcal A_1\mathcal B_2+
\\
&+\mathcal A_2\mathcal B_1 +\mathcal A_2\mathcal B_2
 -2\mathcal A_1-\mathcal A_2-\mathcal B_1\,.
 \end{aligned}
\eeq
Since the maximal value of $\langle I_{3322}\rangle$ with maximally entangled state is 1/4, the Bell parameter in the case of 
detection inefficiencies $\eta_A=\eta_B=\eta$
becomes

\beq
\frac14\eta^2+\frac12\eta(1-\eta)(\mathcal B_1+\mathcal B_2-\mathcal A_1-\mathcal A_2-4)+(1-\eta)^2X_{AB}\,.
\eeq
The choice that minimizes the critical efficiency is given by $\mathcal B_1=\mathcal B_2=1$ and
$\mathcal B_3=\mathcal A_1=\mathcal A_2=\mathcal A_3=0$, giving $X_{AB}=-1$, $T_A=T_B=-\frac12$ and
\beq
\frac14\eta^2-\eta(1-\eta)-(1-\eta)^2>0\,,
\eeq
solved by
\beq
\eta>2(\sqrt2-1)\simeq0.828\,.
\eeq	
The choice of the $\mathcal A$'s and $\mathcal B$'s 
corresponds to choosing for the non-detection events the outcome $-1$ for the inequality written as  $I^{(2)}_{3322}$.

\subsection{Two-qutrit $I_{3}$  inequality}
For this two-qutrit inequality Alice has three outcomes for each  observable $A_j$. 
In the case of non-detection she assigns with probability $\mathcal A^{(1)}_j$ the outcome 1,
 with probability $\mathcal A^{(2)}_j$ the outcome 2, and with probability $1-\mathcal A^{(1)}_j-\mathcal A^{(2)}_j$ the outcome 3.
 The same applies to Bob. For maximally entangled states $p(a_j)=p(b_j)=p(\overline a_j)=p(\overline b_j)=\frac13$, and we have
\beq
\label{newI3}
\begin{aligned}
T_A=&T_B=-\frac23
 \\
X_{AB}=&\mathcal A^{(1)}_1\mathcal B^{(1)}_1+\mathcal A^{(1)}_1\mathcal B^{(1)}_2+\mathcal A^{(1)}_2\mathcal B^{(1)}_1 -\mathcal A^{(1)}_2\mathcal B^{(1)}_2+
\\
&+ \mathcal A^{(2)}_1 \mathcal B^{(2)}_1+\mathcal A^{(2)}_1 \mathcal B^{(2)}_2+\mathcal A^{(2)}_2 \mathcal B^{(2)}_1-\mathcal A^{(2)}_2 \mathcal B^{(2)}_2
\\
&+\mathcal A^{(1)}_1 \mathcal B^{(2)}_1+\mathcal A^{(2)}_1 \mathcal B^{(1)}_2+\mathcal A^{(2)}_2\mathcal B^{(1)}_1-\mathcal A^{(2)}_2\mathcal B^{(1)}_2
\\
&-\mathcal A^{(1)}_1-\mathcal A^{(2)}_1-\mathcal B^{(1)}_1-\mathcal B^{(2)}_1\,.
 \end{aligned}
 \eeq
The optimal choice of $\mathcal A^{(\mu)}_k$'s and $\mathcal B^{(\mu)}_k$'s is the one that maximizes $X_{AB}$.
This term is clearly upper bounded by 0 (it corresponds to the Bell inequality). Then the choice 
$\mathcal A^{(\mu)}_j=\mathcal B^{(\nu)}_k=0$ saturates the bound. 
This choice corresponds to choosing for the non-detection events the outcome 2 for the inequality written as \eqref{I3}.

\end{appendix}
%%%%%%%%%%%%%%%%%%%%%%%%%%%%%%%%%%%%%%%%%%%%%%%%%%%%%%%%%%%%%%%%%%%

%%%%%%%%%%%%%%%%%%%%%%%%%%%%%%%%%%%%%%%%%%%%%%%%%%%%%%%%%%%%%%%%%%%

\end{document}